\documentclass[preprint,preprintnumbers,amsmath,amssymb, floatfix]{revtex4}

\usepackage{graphicx}
\usepackage{dcolumn}
\usepackage{bm}
\usepackage{amssymb}
\usepackage{wasysym}
\usepackage{float}
\usepackage{textcomp}
\usepackage[latin1]{inputenc} 
\usepackage[T1]{fontenc}

\usepackage{tabularx}
    \newcolumntype{Y}{>{\centering\arraybackslash}X}

\begin{document}

\preprint{}

\title{Fractal dimension and size scaling of domains in thin films of multiferroic BiFeO$_{3}$}

\author{G. Catalan$^{1}$}
\email{gcat05@esc.cam.ac.uk}
\author{H. Béa$^{2}$, S. Fusil$^{2}$}
\author{M. Bibes$^{3}$}
\author{P. Paruch$^{4}$}
\author{A. Barthélémy$^{2}$}
\author{J. F. Scott$^{1}$}
\affiliation{$^{1}$Centre for Ferroics, Department of Earth Sciences, University of Cambridge, Cambridge CB2 3EQ, United Kingdom}
\affiliation{$^{2}$Unité Mixte de Physique CNRS/Thales, Route départementale 128, 91767 Palaiseau, France}
\affiliation{$^{3}$Institut d'Electronique Fondamentale, CNRS, Univ. Paris-Sud, 91405 Orsay, France}
\affiliation{$^{4}$Laboratory of Atomic and Solid State Physics, Cornell University, Ithaca, New York, USA}

\date{\today}

\begin{abstract}

We have analyzed the morphology of ferroelectric domains in very thin films of multiferroic BiFeO$_{3}$. Unlike the more common stripe domains observed in thicker films BiFeO$_{3}$ or in other ferroics, the domains tend not to be straight, but irregular in shape, with significant domain wall roughening leading to a fractal dimensionality. Also contrary to what is usually observed in other ferroics, the domain size appears not to scale as the square root of the film thickness. A model is proposed in which the observed  domain size as a function of film thickness can be directly linked to the fractal dimension of the domains. 
\end{abstract}

\maketitle

Multiferroic materials are currently attracting a great deal of attention on account of their interesting physics and potential applications \cite{Fiebig05}. Of these, possibly the most studied is the perovskite BiFeO$_{3}$ (BFO), one of the very few materials which is multiferroic (ferroelectric and antiferromagnetic) at room temperature \cite{Smolenskii, Teague}. Its lead-free nature and large remanent polarization \cite{Wang03} have already motivated Fujitsu to use it as the active layer in prototype ferroelectric memories \cite{Fujitsu}; also, sub-lattice magnetic switching using voltage has been demonstrated \cite{Zhao06}, which may find its way into spintronic applications via exchange bias \cite{Bea06}. The possible coupling between ferroelectric and antiferromagnetic domains has triggered a flurry of work on the morphology and functional properties of the  domains \cite{Zhao06, Zavaliche05, Chu06, Chu07, Chen07}. The ferroelectric domains are generally found to be straight-walled, and to follow the well-known scaling law of Landau, Lifshitz and Kittel (LLK) \cite{Landau35,Kittel46,Kittel49}, that is, their domain size grows proportionally to the square root of film thickness \cite{Chen07}. 

BFO has a complex phase sequence, both in bulk and in thin film form. In bulk, it goes (on cooling) from cubic to orthorhombic to rhombohedral, with a recently-discovered  possible metal-insulator transition at high-temperature \cite{Palai07}. The room-temperature rhombohedral phase is normally monoclinic for epitaxial thin films, but it appears to be tetragonal below a critical thickness, which for BFO grown epitaxially on SrTiO$_{3}$ substrates is of the order of ~100nm \cite{Bea07}. We have analyzed the morphology and scaling of the domains in the small-thickness (tetragonal) regime and found it to be fundamentally different from that observed at higher thickness: i) the domains are not straight, but irregular in shape; ii) there is significant domain wall roughness with a fractal  dimensionality; and iii) the average domain size does not scale as the square root of the film thickness. Taking into account the fractal nature of the domain walls, we have derived a new relation which accounts for the anomalous domain size behaviour, with the scaling exponent being a function of the fractal Hausdorff dimension. 
  
\begin{figure}[h]
\centering
\scalebox{0.78}{\includegraphics{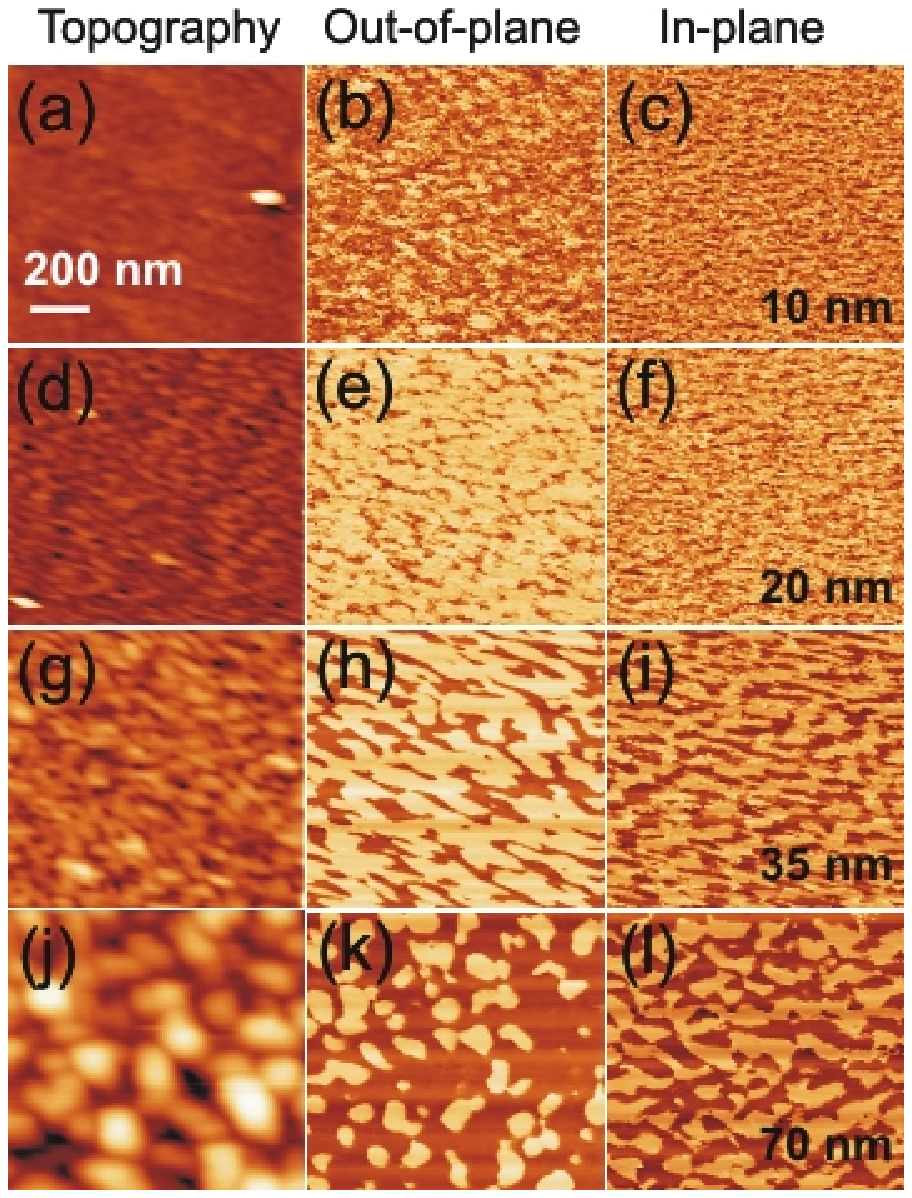}}
\caption{Surface topography and PFM scans of domains in BFO films of different thickness. The PFM cantilever was oriented along a $\langle$100$\rangle$ axis}
\label{Morphology}
\end{figure}

The thin films of BFO were grown by pulsed laser deposition on SrTiO$_{3}$ with a conductive buffer layer of (La,Sr)MnO$_3$ \cite{Bea05}, and the ferroelectric domain morphology was studied by way of piezo-response atomic force microscopy (PFM). In Figure 1 we show the surface topology and PFM response of some films. The shape of these "virgin" domains is highly irregular, and not correlated with topography. Irregularly-shaped domains have also been seen in much thinner films (30nm or less \cite{Chu07}), so it appears to be a general property of ultra-thin BFO, with the critical thickness depending on the choice of substrate and electrode, as well as growth conditions. 

Figure 2 shows the average domain period as a function of film thickness. For comparison, equivalent data for domains in other ferroic systems (ferroelectric or ferromagnetic) are included. While ferroelectric domains are generally smaller than ferromagnetic domains \cite{Hehn96, Streiffer02, Schilling06, Catalan07}, the ferroelectric domains in multiferroic BFO are noticeably bigger than those in "pure" ferroelectrics. This suggests a higher energy cost of the domain walls \cite {Catalan07, Lukyanchuk05}, possibly due to the contribution from the magnetoelectric coupling term. Even more surprising is the fact that the domain periodicity does not scale as the square root of the film thickness. In our case, the exponent for the thickness dependence in $w=Ad^{\gamma}$  (where $w$ is domain period and $d$ is film thickness) is around $\gamma\simeq 3/4$. This is not due to the fact that the domains are of mosaic-type rather than stripes, since such domains are also predicted to follow the $\gamma$=1/2 scaling \cite{Kittel49}, so a different explanation is required. 

\begin{figure}[h]
\centering
\scalebox{0.75}{\includegraphics{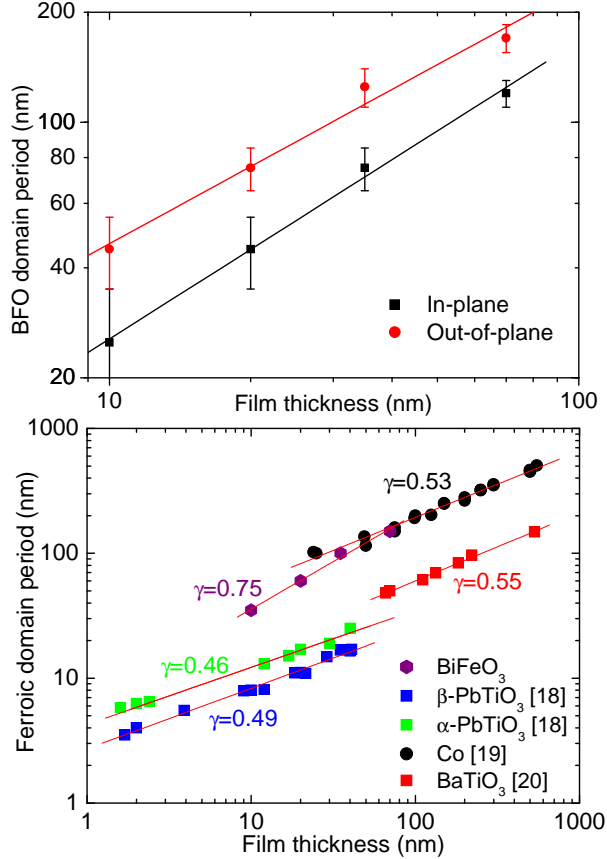}}
\caption{Above, periodicity of in-plane and out-of-plane domain contrast in the PFM scans  as a function of BFO film thickness. The straight lines are least-fit squares giving scaling exponents of 0.70$\pm$0.07 (out-of-plane) and 0.81$\pm$0.04 (in-plane). Below, comparison between the ferroelectric domain periodicity of BFO and the domain periodicity of other ferroics found in the literature.}
\label{Period}
\end{figure}

In order to gain further insight into this problem we have studied the morphology of the domain walls. Firstly, using the program WSxM \cite{nanotec}, we have analyzed the ratio of perimeters ($P$) to areas ($A$) for the spontaneous domains \cite{experimental}. The scaling is such that $P\propto A^{H_{\parallel}/2}$, where $H_{\parallel}$ is the Hausdorff dimension of the domain walls in their projection on the x-y plane accessible by the PFM; if the domain walls were perfectly smooth, $H_{\parallel}$=1, otherwise $1\leq H_{\parallel} \leq 2$. Plotting log($P$) vs log($A$) (figure 3), $H_{\parallel}=1.45\pm 0.2$  was obtained. This value is in close agreement with that found in BFO-doped lead zirconium-titanate (PZT) films, where fractal ferroelectric domains were written using a PFM tip via liquid electrodes \cite{Kalinin07}. 

To relate these observations to the nature of the domain walls, we have also directly measured the domain
wall roughness of AFM-written linear domains, created by alternate applications of negative and positive voltage
($\pm$ 8V) to the film surface. The resulting domain walls were analyzed using the pair-correlation method
described by Paruch et al. \cite{Paruch05}. The correlation function of relative displacements essentially measures the local variance of the domain wall position from an elastically ideal flat configuration as a function of the length L along the wall, and is predicted to show a power-law growth at equilibrium, governed by a characteristic roughness exponent zeta. We observe such a power law growth at short length scales (L<100 nm) followed by a saturation of the correlation function. The value of the roughness exponent $\zeta\sim 0.5-0.6$ is higher than that observed in PZT \cite{Paruch05}, and could be an indication of either a lower dimensionality (being close to the theoretically predicted value of $\zeta$=2/3 for a one-dimensional elastic domain wall in random bond disorder potential), or possibly a different type of disorder playing an important role in a system with combined magnetic and electric effects \cite{dimensionality}.

There are additional self-consistency checks possible for the fractal Hausdorff dimension, requiring time-resolved data for current transients \cite{Ishibashi71}. For well-annealed PZT, analysis of I(t) yielded \cite{Scott88} $H=2.55 \pm 0.25$, compatible with the present results on BFO assuming straight walls normal to the film surface, that is, $H=H_{\parallel}+H_{\perp}$=1.5+1.0 where $H_{\parallel}$ and $H_{\perp}$ are, respectively, the in-plane and out-of-plane components of the fractal dimension. A similar result was also inferred for SBN (strontium barium niobate) by Kleemann \textit{et al.} \cite{Kleeman06}, who observed a dimensionality of $1<H-1<1.7$, inferring $H=2.35\pm 0.35$. The dimensionality of ~2.5 is also compatible with hyperscaling near critical points in ferroelectrics \cite{Scott06, Scott07}. Finally, it is also worth mentioning the work of Yu and Randall \cite{Randall99}, who showed fractal-like (dentritic) domain structures in relaxor ferroelectrics, although the fractal dimension was not calculated in that case. 

The fractal nature of the domain walls, and even the specific Hausdorff dimension of $\sim 2.5$, appear therefore to be rather more general than previously thought, so it is important to understand how fractal walls may affect domain scaling. A fractal distribution of domain sizes (Cantor sets) has been analysed \cite{Ozaki93} for ferroelectric stripe domains, but the walls in these studies were straight, and a conventional LLK exponent $\gamma=1/2$ was still found, unlike that of our films. Here, instead, we have modified the domain size scaling problem in order to incorporate the fractal roughening. The LLK law arises from the need to minimize the total energy of the domains and that of the domain walls. The energy density of the \textit{domains} is proportional to their size $w$ (irrespective of whether they are stripe-type or mosaic-type \cite{Kittel49}):

\begin{equation}
E_{domain}=Uw	
\label{DomainEnergy}						
\end{equation}

Where $U$ is a constant arising from either depolarization, demagnetization, strain, or a combination of them. The energy of the \textit{domain walls} is proportional to their number density, which is found dividing the area of the film by the area of each domain, i.e., $N=L^{2}/w^{2}$=number of domain walls. The energy of each domain wall is equal to its energy density ($\sigma$) times the area of the wall, which is essentially the domain perimeter times the thickness of the film: $\sigma Pd$. If the domain walls are fractal, then the domain perimeter does not scale linearly with the average domain size $w$, but fractally as $P=w^{H_{\parallel}}$, where $H_{\parallel}$ is the Hausdorff dimension. Hence $E_{wall}=\sigma w^{H_{\parallel}}d$. The energy of all the domain walls in the crystal is thus $E_{wall}\times N=\sigma w^{H_{\parallel}}d L^{2}/w^{2}$, and the energy density (per unit area of film) is found dividing by $L^{2}$. Thus:

\begin{equation}
E_{walls}=\sigma w^{H_{\parallel}}\frac{d}{w^{2}}					
\label{WallEnergy}
\end{equation}

The optimum domain size is found by adding (1) and (2) and minimizing with respect to $w$. This leads to

\begin{equation}
w=\left[\left(2-H_{\parallel}\right)\frac{\sigma}{U}\right]
^{\left(\frac{1}{3-H_{\parallel}}\right)}d^{\left(\frac{1}{3-H_{\parallel}}\right)}=kd^{\left(\frac{1}{3-H_{\parallel}}\right)}						
\label{DomainPeriod}
\end{equation}

The above treatment assumes that the domain walls are straight in the vertical direction. Otherwise, the vertical size of the domain walls would scale as $d^{H_{\perp}}$, and Eq. \ref{DomainPeriod} would become instead:
 
\begin{equation}
w=kd^{\left(\frac{H_{\perp}}{3-H_{\parallel}}\right)}						
\label{DomainPeriod3d}
\end{equation}
For $H_{\perp}$=$H_{\parallel}$=1 (flat walls) we recover the standard Kittel result $\gamma=1/2$. 

PFM does not allow establishing whether the walls are rough or straight along the z-direction. Previous work in PZT \cite{Paruch05} and Bi$_4$Ti$_3$O$_{12}$ \cite{Katayama07} shows that domain walls may be rough along one direction and flat along another, due to anisotropy in the strength of the dipole-dipole and elastic interactions. Here, comparison between eq. \ref{DomainPeriod}, eq. \ref{DomainPeriod3d}, and the measured domain periodicity as a function of thickness can help to indirectly determine $H_{\perp}$. 

\begin{figure}[h]
\scalebox{0.78}{\includegraphics{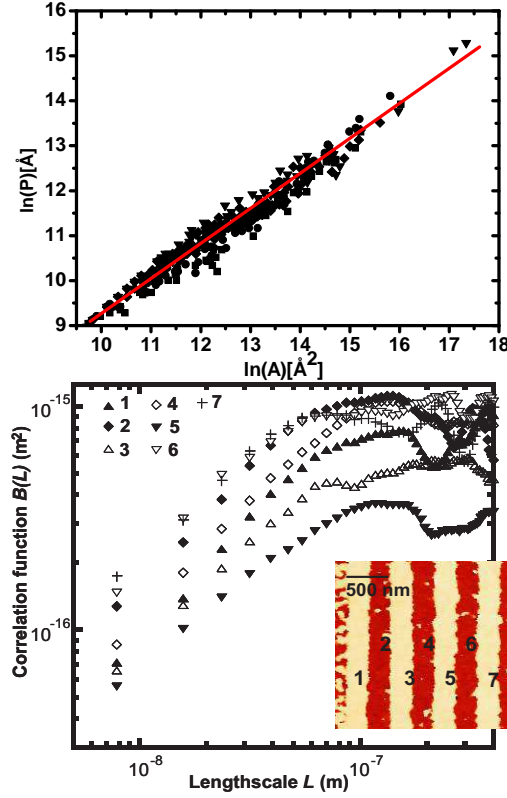}}
\caption{Above, perimeter as a function of area for spontaneously formed domains. The slope of the log-log plot is related to the Hausdorff dimension: $P\propto A^{H_{\parallel}/2}$. Below, correlation function for the roughness of the written domain walls as a function of lengthscale, measured for a 70 nm thick BFO film.  An average value of the roughness exponent exponent $\zeta$=0.56 is obtained for the seven domain walls indicated. This higher value of $\zeta$ for BFO compared to that for PZT suggests either a lower dimensionality in a random bond pinning scenario, or the presence of stronger individual pinning centers which could also be responsible for the fractal structure of the spontaneous domains in the BFO films.}
\label{Roughness}

\end{figure}

For the measured $H_{\parallel}\simeq$ 3/2, assuming straight walls along the vertical direction, eq. \ref{DomainPeriod} predicts a value  $\gamma\simeq$2/3 for the exponent in $w=kd^{\gamma}$. If, on the other hand we assume that $H_{\parallel}$=$H_{\perp}$, then eq. \ref{DomainPeriod3d} predicts $\gamma$=1 instead (although it is worth mentioning that values of $\gamma$ approaching 1 would also be consistent with the classical theory for ferroelastic domains, which predicts  $\gamma$=1 for small thickness \cite{Pertsev95}). The actual measured value for the periodicity of the in-plane domains is (see Figure 2) $\gamma\approx$0.7-0.8, which suggests that the domain walls are anisotropically roughened, being more irregular in the horizontal direction than in the vertical one.  

The present theoretical framework is valid not just for BFO, but for any ferroic (ferroelectric, ferromagnetic or ferroelastic) material with fractal walls. Although it is consistent with the empirical results, the domains in the thinnest films are close to the resolution limit of PFM, and thus there is a substantial error bar associated with their measured periodicity. Also, the value of the scaling exponent $\gamma$ appears to be different for out-of-plane and in-plane variants (Figure 2-a), although the difference is within the statistical error. Finally, while the fractal analysis was performed on the samples with best signal-to-noise ratio, there may still be some electronic noise contribution to the fractal dimension.

In summary, we have analyzed the domain periodicity and domain morphology of very thin films of BFO and found: i) the spontaneous ferroelectric domains in BFO are generally bigger than those in other ferroelectrics, consistent with an increased energy cost of the domain walls due to magnetoelectric coupling; ii) the domains are irregularly shaped and characterized by a fractal dimension; iii) the domain size does not scale as the square root of the film thickness; and iv) the unusual  scaling can be directly related to the the fractal dimension of the walls. Specific further tests on the fractal dimensionality inferred are suggested, particularly the use of switching current transients. We hope that these findings will motivate more research into the physics of domain walls in ferroic and multiferroic systems.  

We acknowledge financial support from ANR-FEMMES and from a Marie Curie Fellowship (GC).


\begin{references}
\bibitem{experimental}In order to minimize spurious contributions from electronic noise to the apparent wall roughening, we analysed only the samples with the biggest signal-to-noise ratio (35 and 70nm), and then only those domains with an area bigger than 6 pixels.
\bibitem{dimensionality}For a random field system $\zeta=(4-d_{eff})/3$, giving a domain wall dimensionality of 2.5 for the observed values of zeta. However, we note that studies in both pure ferromagnetic and ferroelectric materials both showed random bond rather than random field disorder, making the random field scenario less likely. It is also worth emphasizing that the non-integer effective dimensions extracted from the roughness analysis do not imply a real fractal structure of the walls, although a fractal structure would give non-integer dimensions.
\end{references}
\end{document}